# Nonwoven scaffolds for bone regeneration

Elaine R. Durham, Giuseppe Tronci, Xuebin Yang, David J. Wood, Stephen J. Russell

University of Leeds, United Kingdom

## 1. The structure of bone and the mechanisms for self-repair

Bone is one of the most commonly transplanted tissues, with 2.2 million bone grafts performed annually worldwide (Tronci et al., 2013a). Surgeons face a diverse spectrum of clinical challenges in bone reconstruction, and this diversity reflects the variety of anatomic sites, defect sizes, mechanical stresses, and available soft tissue cover. Autologous bone grafting remains the gold standard for the reconstruction of skeletal defects (McMahon et al., 2013), although drawbacks, including limited supply, bone graft loss/resorption, and donor site morbidity, impose a pressing demand for advanced biomaterial solutions. For these reasons, the World Health Organization (WHO) has confirmed the current decade as the 'Bone and Joint Decade'.

To develop successful bone scaffolds, clinicians must adopt a multidisciplinary approach in order to understand and stimulate the natural bone regeneration process. In addition to an understanding of cell biology and genetics, this approach requires knowledge of bone structure and its hierarchical organisation, from the macro (centimetre) to the nano (extracellular matrix, ECM) scales (Figure 1). At the macroscopic structural level, bone consists of a dense shell of cortical bone that supports and protects. The interior porous cancellous bone optimises weight transfer and minimises friction at the articulating joints (McMahon et al., 2013). Cortical bone is composed of repeating osteon units, whereas the cancellous bone is made of an interconnecting framework of trabeculae with bone marrow-filled free spaces. These trabeculae and osteon units are composed of collagen fibres. In the osteons, 20–30 concentric layers of fibres, called lamellae, are arranged at 45° surrounding the central canal, and they contain blood vessels and nerves.

Moving from the macroscopic to the molecular level, bone is a composite material, consisting of cells embedded in the ECM. The ECM plays a key role in the localisation and presentation of biomolecular signals, which are vital for neo-tissue morphogenesis. Proteoglycans, glycosaminoglycans, and mineralised collagen are integrated in a supramolecular hydrogel. Collagen fibrils are arranged with a 67-nm periodicity and 40 nm gaps where hydroxyapatite crystals are situated. The mineral phase is thought to dominate the stiffness of bone, which increases more than linearly with mineral content. The toughness of the material is thought to mainly arise from its hierarchical organisation, whereby the lowest hierarchical level contributes to the outstanding bone fracture resistance (Dunlop and Fratzl, 2010). This unique hierarchical organi- sation enables bone to exhibit mechanical properties far superior to those of its single components. Thus, the repair and reconstruction of bone defects require innovative strategies that can closely mimic the complex tissue hierarchical organisation.

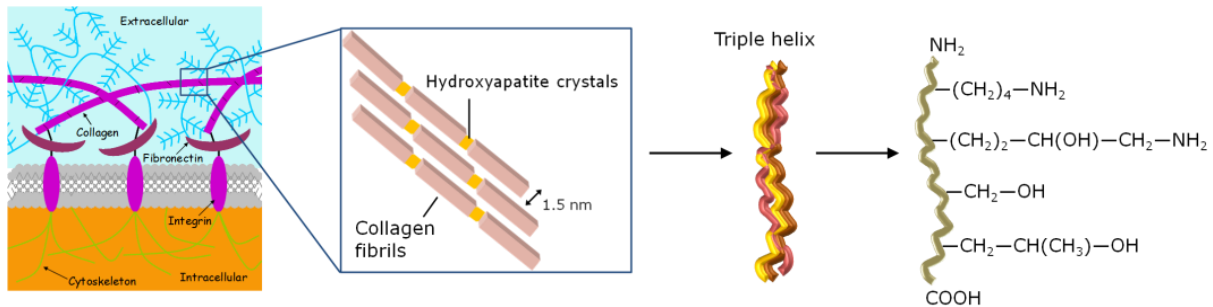

**Figure 1.** Hierarchical organisation of bone over different length scales. B one tissue structure, function, and shape are regulated by the extracellular matrix (ECM), a supramolecular hydrogel network, in which cells are immersed. The ECM is mainly composed of collagen fibrils mineralised with hydroxylapatite crystals. Fibrils result from the assembly of collagen triple helices, which are based on left-handed polyproline chains at the molecular level.

For bone regeneration, bone healing processes must also be given careful consideration. Bone has the capacity to repair itself, and this self-repair can be harnessed to repair small bone defects, heal non-unions, and lengthen short bones. Fracture healing is a complex regenerative process initiated in response to injury, in which bone can heal by primary or secondary mechanisms (Alman et al., 2011). In primary healing, new bone is laid down without any intermediate. This type of healing is rare in a complete bone fracture, except when the fracture is rigidly fixed through certain types of surgery. In the more common secondary mechanism of healing, immature and disorganised bone (i.e., callus) forms between the fragments. During the fracture repair process, cells progress through stages of differentiation reminiscent of those that cells progress through during normal foetal bone development. In the normal development of long bone, undifferentiated mesenchymal cells initially form a template for the bone, which then differentiates into chondrocytes. After this phase, blood vessels enter the cartilaginous template, and osteoblasts, which differentiate from perivascular and other cells surrounding the bone, form bone. The reparative process is impaired in large, critical-sized bone defects (i.e., gaps beyond 2.5 times the bone radius) (Schroeder and Mosheiff, 2011), and osteoblastic differentiation is inhibited, with undifferentiated mesenchymal tissue remaining at the fracture site. Such defects can be caused by blunt or penetrating trauma, surgical treatment of tumours or necrosis caused by radiation, or various chemical substances. These defects represent a considerable surgical challenge, are associated with high socioeconomic costs, and highly influence patients' quality of life, both private and professional (Woodruff et al., 2012). Despite huge progress being made toward the design of bone implants, the integration of all tissue properties and functions in a single biomaterial system remains a major research challenge. Researchers still need to develop reliable tools enabling them to systematically and temporally control the material structure and organisation, properties and functions, so that next-generation scaffolds can successfully ensure full clinical relevance. Fibrous assemblies in the form of nonwoven scaffolds have been repeatedly investigated over the last 20 years in relation to tissue regeneration. As the requirements of clinicians become more and more tissue specific, and synthetic biofunctional biomaterials are developed, nonwoven scaffold engineers must be able to respond with the capability to produce truly biomimetic architectures.

## 2. Fibre manufacture from biomaterials

Biomaterials for bone regeneration should be biocompatible, biodegradable substances that support cell attachment, spread, proliferate (osteoconductive), and control cell differentiation into osteogenic lineages (Dawson et al., 2011; El-Gendy et al., 2013; Yang et al., 2003b). For a biomaterial to be suitable for manufacture into a nonwoven scaffold, it must be suitable for conversion into fibres, and the method of extrusion should not adversely affect biocompatibility or the properties of the material. A variety of natural and synthetic biomaterials can be converted into fibres, with different extrusion methods being applied depending on the composition of the biomaterial. These include, wet and dry solution spinning, melt spinning, and gel spinning. Materials extracted from a natural source, such as collagen, are attractive biomaterials, but the retention of the native structure after solubilisation and the precipitation of the regenerated product after spinning remain challenging. Synthetic biomaterials, notably those exhibiting thermoplastic behaviour, are often more straightforward to extrude due to their excellent mechanical properties, but such materials can lack suitable surface chemistry for cell attachment, and they might produce undesirable degradation products in vivo. One issue can be the inflammatory response of the surrounding tissues while the polymer hydrolytically degrades (Agrawal and Ray, 2001; Cai et al., 2007). Hybrid materials containing both synthetic and natural materials, such as regenerated collagen or blends of two or more polymers, provide one means for balancing the required mechanical and chemical prop- erties. Additionally, the incorporation of materials such as hydroxyapatite tricalcium phosphate (Day et al., 2005; Mastrogiacomo et al., 2006) during spinning also enables the properties of the bulk product to be substantially modified according to specific clinical requirements. Both the biomaterial composition and the method of fibre extru- sion affect the bulk properties of the resulting nonwoven scaffold architecture. One scaf- fold parameter that has received considerable attention is the production of fibres of submicron diameter. Nanofibrous scaffolds have long been championed by the tissue engineering community with early work on electrospun polylactide nanofibre-based tissue engineering scaffolds indicating that human mesenchymal stem cells tended to proliferate more on nanofibre scaffolds than on microdiameter fibrous scaffolds (Shanmugasundaram et al., 2004). Since then, many reports have shown that nanofi- brous scaffolds support cell growth for tissue regeneration (Chung et al., 2011; Huang et al., 2011; Kumbar et al., 2008). However, conflicting evidence also suggests that nanofibres are only preferential to microfibres when using synthetic biomaterials (Rnjak-Kovacina and Weiss, 2011). One of the challenges in designing three-dimensional scaffolds containing submicron fibres that pack closely together is facilitating adequate cell penetration into the full thickness of the fibrous assembly. The selection of an appropriate biomaterial from which to manufacture a nonwoven scaffold is, of course, a major consideration, and in bone regeneration, the evaluation of mate- rials has included regenerated collagen, gelatin (Venugopal et al., 2008), poly(lactic acid) (Kim et al., 2006), poly(lactic-co-glycolic acid) (Liao et al., 2008), silk fibroin (Kim et al., 2005), chitosan (Shin et al., 2005), and poly(ε-caprolactone) (Porter et al., 2009). For brevity, further discussion in this chapter is restricted to collagen and poly(ε-caprolactone) nonwoven production.

## 2.1 Collagen

Collagen has been widely applied for the design of tissue-like matrices for repair (Hong et al., 2010) because of its natural occurrence in bone tissue. The collagen molecule is based on three left-handed polyproline chains, each of which contains the repeating unit Gly-X-Y, with X and Y being predominantly proline (Pro) and hydroxyproline (Hyp), respectively. The three chains are staggered from one another by one amino acid residue and are twisted together to form a right-handed triple helix (300 nm in length, 1.5 nm in diameter). In vivo, triple helices can aggregate to form collagen fibrils, fibres, and fascicles, which are stabilised via intermolecular enzymatic crosslinking (Buehler, 2008; Grant et al., 2009). However, collagen properties are challenging to control in physio- logical conditions due to the fact that collagen's unique hierarchical organisation and chemical composition in vivo can only be partially reproduced in vitro. As a result, regenerated collagen materials display non-controllable swelling and weak mechanical properties in physiological conditions. The limited solubility in organic solvents, high swelling in aqueous solution, and non-controllable batch-to-batch variation in chemical properties represent significant challenges to the use of collagen for the design of defined, water-stable nonwovens. To improve its stability, collagen has been widely crosslinked with N-(3-dimethylaminopropyl)-N'-ethylcarbodiimide hydrochloride (EDC) (Olde Damink et al., 1996), glutaraldehyde (GTA) (Olde Damink et al., 1995b), and hexamethylene diisocyanate (HDI) (Olde Damink et al., 1995a). In the first case, zero-length covalent net-points are formed so that no harmful and potentially cytotoxic molecules are introduced (Haugh et al., 2011). Due to the minimal net-point length, however, crosslinking of adjacent collagen molecules is unlikely because the terminal amino functions are too far apart to be bridged, resulting in non-varied mechanical properties. In contrast to EDC, GTA and HDI involve the covalent incorporation of oligomeric segments between distant collagen molecules. The reaction of collagen with aldehydes or isocyanates in aqueous solution has been reported to result in a cascade of non-controllable side reactions and the formation of highly reactive and potentially toxic functional groups coupled to the polymer backbone (Zhang et al., 2011). To avoid such undesirable side reactions, collagen has been crosslinked with diimidoesters, such as dimethyl suberimidate, 3,3'-dithiobispropionimidate, and acyl azide, resulting in stable materials in physiological conditions, although the extensibility of the material was reduced (Charulatha and Rajaram, 2003). Dehydrothermal treatment or riboflavin-mediated photocrosslinking has also been applied as physical, benign crosslinking methods, although partial loss of native collagen structure and nonhomogeneous cross- linking was observed in these cases (Weadock et al., 1995). Rather than direct covalent crosslinking, alternative approaches have recently focused on the formation of injectable ECM-mimicking gels via synthetic collagen blends (Hartwell et al., 2011), as well as the design of cell-populated matrices via derivatisation with cinnamate (Dong et al., 2005) or acrylate (Brinkman et al., 2003) moieties. In these methods, although the resulting mechanical properties may be enhanced, synthetic components, such as polymers or co-monomers, are required to promote the formation of water-stable matrices, and the alteration of the protein backbone and biofunctionality may result.

Reliable synthetic methods must therefore be applied in order to improve thermo-mechanical behaviour, without affecting biofunctionality (de Moraes et al., 2012), specifically biocompatibility and bioactivity. To address these challenges, collagen fibres can be chemically functionalised (Tronci et al., 2013c), so that a covalent net- work is established at the molecular level (Tronci et al., 2013b); in this way, temporal stability of both fibres and mesh architecture may be ensured in physiological conditions.

Collagen fibrillogenesis can be induced in vitro by exposing monomeric collagen solutions to physiological conditions, resulting in viscoelastic gels at the macroscopic level (Lai et al., 2011). The design of collagen mimetic peptides has also been proposed as an alternative strategy for recapitulating the multiscale organisation of natural collagen (O'Leary et al., 2011). However, despite the formation of hierarchical triple helix assemblies, the resultant thermal and mechanical stability is still not adequate for biomaterial applications, so that chemical functionalisation of side- or end-groups is crucial.

The functionalisation of collagen requires careful consideration, because the hier-archical organisation of collagen imposes constraints in terms of protein solubility, the occurrence of functional groups available for chemical functionalisation, and material biofunctionality. As improved synthetic methods are developed, functionalised collagen with preserved protein conformation and full biocompatibility will become available, enabling the manufacture of better performing biomimetic, nonwoven architectures. There have been two distinct approaches to collagen fibre formation. The tissue engineering community has mainly focused on electrospinning, whereas industries producing artificial hair and textile fibres have exploited conventional wet-spinning approaches.

Electrospinning of collagen involves the extrusion of a positively charged polymer solution with the resulting fibres being collected as a nonwoven web on a grounded or negatively charged collector. The majority of studies involving the production of collagen fibres without a second carrier polymer involve the use of 1,1,1,3,3,3-hexafluoro-2-propanol (HFIP) as the solvent. The popularity of HFIP is based on the dual role it plays in collagen solubilisation – two trifluoromethyl groups serve to break hydrophobic interactions, and the mildly acidic secondary alcohol hydroxyl group assists in breaking the hydrogen bonds (Dong et al., 2009). However, cytotoxicity and the destructive effect on the structure of functionalised collagen have inspired a search for more benign sol- vents. Both collagen and gelatine have been successfully electrospun in a mixture of phosphate buffer saline/ethanol to produce fibres with average diameters ranging from 0.21 to 0.54 mm, depending upon the salt concentration (Zha et al., 2012). Note that vacuum drying is still normally required to remove the solvent after spinning.

The traditional wet spinning of collagen fibres has two major advantages. First, fibres can be produced with larger diameters, if required, and the fibre orientation and architecture of the nonwoven scaffold can be manipulated independent of the spinning process. Second, the stretching and drying of the extruded filaments are easier to control. In its simplest form, wet spinning involves dissolving the polymer in an appropriate solvent and then extruding the polymer solution via a spinneret into a coagulation bath containing a non-solvent. Collagen has been solubilised in acidic environments and wet-spun into coagulation baths containing different ethanol/ acetone mixtures, resulting in fibres with diameters ranging from 89 to 140 mm and tenacities of between 8.5 and 8.3 cN/tex (Meyer et al., 2010). Such fibre dimensions are larger than those typically targeted in the manufacture of tissue engineering scaffolds, but smaller diameters may be obtained through the appropriate control of manufacturing conditions. Owing to the excellent control of fibre dimensions and the high delivery speeds that are possible during production, wet spinning is an attractive route for the production of functionalised collagen fibres. Melt or thermo- plastic spinning of collagen to produce fibres has also been reported, but the high temperatures required cause denaturing and are likely to disrupt functionalisation, so resulting materials are unlikely to be appropriate for tissue engineering applications (Meyer et al., 2010).

The structural and mechanical properties of chemically functionalised collagen fibres are of paramount importance in terms of the biocompatibility and mechanical performance for bone tissue engineering. Two major challenges in the production of collagen using synthetic methods are the ability to control material stability in physiological conditions and the preservation of the native protein conformation. Preservation of the triple helix structure is crucial for ensuring enzymatic implant degradability. Here, intact triple helices are required to promote effective degradation of fibrillar collagen (Gaudet and Shreiber, 2012). If the tertiary structure is significantly altered, then collagenase degradation rates could also be affected, in turn influencing material biodegradability, as well as the extent to which cells remodel the scaffold. Attenuated total reflectance and Fourier transform infrared spectroscopy (ATR-FTIR) is a useful analytical technique for elucidating the protein molecular conformation in collagen scaffolds manufactured using a synthetic route. Collagen displays distinct amide bands via FTIR, which characterise its triple helix structure. These are (i) amide A and B bands at 3300 and 3087 $cm^{-1}$, respectively, which are mainly associated with the stretching vibrations of N–H groups; (ii) amide I and II bands, at 1650 and 1550 cm$^{-1}$, resulting from the stretching vibrations of peptide C=O groups as well as from N–H bending and C–N stretching vibrations, respectively; (iii) an amide III band centered at 1240 cm$^{-1}$, assigned to the C–N stretching and N–H bending vibrations from amide linkages, as well as wagging vibrations of CH$_2$ groups in the glycine backbone and proline side chains. Each of the previously mentioned amide bands should be exhibited in the FTIR spectra of functionalised collagen, whereby no detectable band shift will be displayed compared to the spectrum of native type I collagen (Figure 1). Other than qualitative findings on unchanged band positions, the FTIR absorption ratio of amide III to the 1450 cm$^{-1}$ band ($A_{III}/A_{1450}$) is usually determined in order to quantify the degree of triple helix preservation. In such a case, an amide ratio close to unity is associated with the preserved integrity of the triple helices (He et al., 2011) following functionalisation of native collagen (Tronci et al., 2013b).

Other valuable analytical techniques for elucidating protein backbone conformation are circular dichroism (CD) and sodium dodecyl sulphate-polyacrylamide gel electrophoresis. CD is based on the fact that single collagen polyproline chains are stabilised into a triple helix structure via hydrogen bonds oriented perpendicularly to the triple helix axis, resulting in an optically active protein (Djabourov, 1988). This molecular feature is exploited by CD spectroscopy to investigate any alteration in protein conformation following either chemical functionalisation or fibre formation. Resulting collagen-based materials are dissolved in dilute acidic conditions and excited with plane-polarised light. Plane-polarised light can be viewed as being made up of two circularly polarised components of equal magnitude, one rotating counter- clockwise (left-handed, L) and the other clockwise (right-handed, R). CD refers to the differential absorption of these two components. If, after passage through the sample being examined, the L and R components are not absorbed or are absorbed to equal extents, the recombination of L and R would regenerate radiation polarised in the original plane. However, if L and R are absorbed to different extents, the resulting radiation would be said to possess elliptical polarisation, resulting in a CD signal (Kelly et al., 2005). One of the most significant advances that has been made in relation to scaffold performance and which will aid future work on the development of functionalised collagen scaffolds is the ability to monitor scaffolds in vivo (Cunha-Reis et al., 2013).

## 2.2 Poly(ε-caprolactone)

Poly(ε-caprolactone) (PCL) is a biocompatible and bioresorbable semicrystalline aliphatic polyester that has been extensively reported in connection with medical applications including bone repair and regeneration. PCL has a degradation time between 2 and 4 years; however, the degradation time can be greatly altered by blending it with another polymer such as collagen. Owing to its biodegradability and exceptional mechanical properties, PCL fibre has been studied extensively in relation to bone engineering (Porter et al., 2009). Its popularity also results from its relatively low cost, ease of processing, and compatibility with both melt spinning, wet spinning, and electrospinning. The major disadvantage of PCL as a biomaterial is its lack of biofunctionality, although this can be compensated for, to some extent, by blending it with other biofunctional materials.

The solvent electrospinning of PCL is well documented, with a large number of studies reporting PCL fibre spinning using the chloroform:methanol solvent system. By adjusting the ratio of chloroform to methanol, applied voltage, and tip to collector distance, electrospun webs with an average fibre diameter ranging from 2.3 to 10.8 mm can be obtained (Pham et al., 2006). Submicron PCL fibres can also be successfully spun using a solvent mixture of chloroform: N,N-dimethylformamide (Pham et al., 2006), and there is potential to produce webs in which both PCL nanofibres and micro- fibres are present, using the same equipment, although producing fibre diameters that are considerably larger than 10 mm remains a limiting factor. Melt electrospinning uses a similar set-up, but, because the polymer is extruded in a molten state, a heating element is required. In this process, the upper limit of the average fibre diameter that can be produced increases between 6 and 30 mm (Detta et al., 2010). The high temperatures that are required during melt extrusion processes means that collagen and other biofunctional materials cannot be easily incorporated in the manufacture of fibres without the risk of degradation, however. Although melt spinning using operating temperatures of 85–90 °C is not appropriate for the production of PCL/collagen blends, biofunctional materials can be incorporated post-spinning in the form of coatings on the fibre surfaces. Traditional melt spinning (as opposed to melt electrospinning) enables the production of large quantities of fibre (Charuchinda et al., 2003) and the ability to produce either continuous filament or staple fibre. This provides greater versatility in the available nonwoven production routes and therefore the types of scaffold architecture and physical properties that can be manufactured.

Wet-spun PCL fibres can be extruded from polymer solution containing acetone and then spun directly into a methanol coagulation bath. As-spun PCL fibres have been reported with a diameter of 150 mm, reduced to 67 mm by cold drawing using an extension of 500% (Williamson et al., 2006). A major benefit of wet spinning, as opposed to the melt-spinning route, is the potential to produce mixed polymer fibres that incorporate the mechanical properties of PCL with the biofunctionality of a material such as collagen. However, appropriate co-solvents are required to facilitate this. The majority of studies have reported the use of HFIP as a solvent system for electrospinning blends of collagen and PCL. This is less than ideal because of the cytotoxic nature of any residual HFIP present in the as-spun fibres. Some alternative approaches have been developed, including those reported by Chakrapani et al. (2012) who used acetic acid as a solvent system for PCL and collagen mixtures.

To improve the osteoconductivity and mechanical properties of scaffolds for bone engineering and to create a pH buffer against the acidic degradation of synthetic polymer matrices, researchers have investigated the incorporation of HA particles (Puppi et al., 2011). Ji et al. (2012) reported that nano-HA platelets significantly improved the mechanical properties, including the strength, strain, and toughness of electrospun collagen fibres. Puppi et al. (2011) wet-spun PCL fibres containing HA using acetone as a solvent and ethanol as a nonsolvent, and they produced fibres with diameters in the range of 100–250 mm.

## 3. Design and assembly of scaffold architectures

In scaffold-guided tissue regeneration, three-dimensional scaffolds serve as temporary tissue substitutes that promote tissue regeneration at the defect site (Langer and Vacanti, 1993). Scaffold design has to take into account the macroscopic properties of the material (e.g., degradability, porosity, and mechanical properties), processability, and the interaction with cells. Suitable porosity and pore interconnectivity are necessary to promote cell migration and differentiation, including within the interior of the scaffold; diffusion of oxygen and nutrient to cells (Botchwey et al., 2003); and removal of waste products from the scaffold. The same structure must also have mechanical properties that enable sharing of the physiological load, if functional neo-tissue is to develop.

In a clinical setting, a nonwoven scaffold can be used alone or in combination with other materials, growth factors, and/or cells. If the scaffold is to be used alone, it must be designed to act as a supporting structure that recruits stem/stromal cells from sur- rounding tissues (Shi et al., 2013; Zhang et al., 2008). In other instances, the scaffold can be designed and used as a carrier vehicle for bioactive growth factors in order to enhance osteoinduction, chondroinduction, and angiogenesis that are crucial for bone and osteochondral-tissue engineering (Green et al., 2004; Yang et al., 2004). Nonwoven scaffolds can also be used in combination with stem/stromal cells that can be directly delivered into the bone defect area to provide both osteogenesis and support elements for bone tissue engineering (Udehiya et al., 2013; Yang et al., 2001). Scaffolds may be bioactive and contain autogenic and/or allogenic stem/stromal cells that can be directly delivered into the bone defect area to provide osteogenesis, osteoinductive, and osteoconductive elements for bone tissue engineering (Yang et al., 2003a,b, 2004). It is therefore important that the required function of the nonwoven scaffold and the selection of materials are carefully considered during the design and development process.

Generally, nonwoven fabrics are highly porous, low-density fibrous assemblies of typically less than 0.40 g/cm$^3$. The internal pore structure is highly interconnected, and there is a relatively wide pore size distribution that can be manipulated during nonwoven fabric manufacturing. The majority of fibres in nonwoven structures are arranged in a planar, x–y orientation, with only a limited number of processes, notably air-laid, vertically lapped, carded webs, with needling and hydroentangling capable of producing a degree of fibre orientation through-thickness. The fibre orientation distribution, which can be manipulated during production of the nonwoven scaffold, strongly influences the isotropy of fabric properties including directional mechanical properties and fluid transport.

Different nonwoven architectures can be produced in the form of a scaffold depending upon the selection of manufacturing route. Nonwoven manufacture typically involves at least two sequential steps: web formation and bonding. Both dry-laid (e.g., carded, carded and lapped, air-laid) and wet-laid web formation processes utilise staple fibres that are cut to a predetermined length prior to nonwoven fabric manufacture. In contrast, spun-melt (e.g., spun-bond, melt-blown) and other direct filament deposition techniques, such as electrospun and force-spun web formation techniques, rely on the direct collection of continuous filaments in the form of a web. Depending on the polymer composition of the fibres in the web, one or more bonding techniques may be applied: mechanical (needling, hydroentangling, stitch-bonding), thermal bonding, or chemical bonding. Mechanical bonding relies on increasing fibre entanglement and frictional resistance within the web to increase resistance to slippage, and therefore, fibre length, fibre diameter, breaking elongation, flexural rigidity, fibre mobility, and fibre–fibre friction are influential parameters.

In dry-laid web formation, the carding of fibres to produce a web involves their progressive disentanglement as they are carried on rollers clothed in wire teeth, which involves fibre–metal and fibre–fibre friction. The low melting point of some biomaterials such as PCL (e.g., 60 °C) can therefore give rise to unwanted fusing of fibres during the carding as a result of frictional heating. Another challenge is the potential for excessive fibre breakage in biomaterial fibres that have low breaking extensions of less than 5%. Short-cut (<15 mm length) biomaterial fibres can also be air-laid to form webs, minimising the potential for fibre breakage during web formation, but mechanical bonding normally results in relatively weak scaffolds. An advantage of wet-laid processes is the high weight uniformity that can be achieved in the web; how- ever, the process requires suspension of short cut fibres in a liquid medium, which can be unsuitable for biomaterials that have poor aqueous stability.

Thermal bonding of synthetic thermoplastic biomaterials such as PCL is also feasible. Practically, fibres with a concentric core-sheath bicomponent structure are preferred in order to minimise thermal shrinkage during heating and to maximise fabric strength while preserving the required internal fabric structure. Owing to biocompatibility issues, chemical bonding, in which an adhesive is applied to the web to prevent fibre slippage, is not normally considered appropriate for tissue scaffold manufacture, unless the binder is, in itself, a biomaterial suitable for invasive use.

The production of nonwoven scaffolds with reproducible architectural features is challenging and remains an important issue with respect to quality control. In mechanically bonded tissue scaffolds, methods of controlling certain features of the internal architecture, such as pore structure, have been attempted by utilising templates (Durham et al., 2012). An example of a highly porous nonwoven architecture made by thermally bonding a carded web of bicomponent poly(lactic acid) (PLA) fibres around a removable spacer template to tune the internal pore structure is shown in Figure 2. Such modifications to existing nonwoven processes highlight the potential for manipulating scaffold structures in a reproducible manner during their production.

In relation to biomimetic scaffolds, a major advantage of nonwovens is their highly interconnected pore structure and the scope that is available to control pore size distribution during manufacturing. The production of scaffolds with appropriate pore sizes is fundamental for their functionality. The minimum pore size for bone engineering is approximately 100 nm, due to cell size, migration requirements, and nutrient transport (Hutmacher et al., 2007). Mean pore sizes in electrospun webs are normally substantially lower than 100 nm, which means initial cell penetration into a thick, three-dimensional scaffold can be impeded. Increasing the fibre diameter is one approach to making larger pores, but this is not always practicable, depending on the biomaterial and spinning conditions. Other methods for increasing pore size involve combining electrospun polymers with sacrificial material, such as a porogen, that can be subsequently removed (Nam et al., 2007; Phipps et al., 2011). Force-spinning, which relies on using centrifugal forces instead of electrical charge to stretch the polymer jet (McEachin and Lozano, 2012), is an interesting candidate for scaffold production because the submicron diameter fibres that are deposited on the collector are less densely packed than they are in electrospinning.

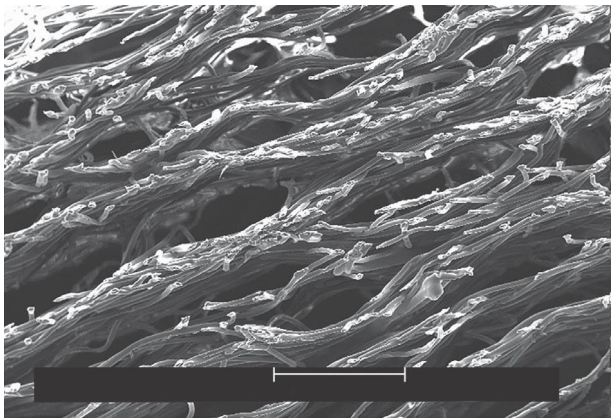

**Figure 2.** Cross section of a nonwoven scaffold produced by carding and through-air thermal bonding (100% core-sheath PLA staple fibre). The cavities correspond to regions where solid metallic spacer elements inserted prior to thermal bonding have been removed.

Given that physiological tissue, including bone, is heterogeneous in structure, the ability to manufacture templated features in electrospun webs that biomimic the native structure is of interest to tissue engineers. Templating techniques have included use of an electronic circuit chip as a collector, which enabled the production of features on the order of several hundred microns (Wu et al., 2010). The majority of other approaches have used patterns in the range of millimetres. Producing well-defined features at the micron level may be hampered by electrical charge jumping across small nonconductive areas, however. Vaquette and Cooper-White (2011) investigated 'round' collectors with a 1-mm diameter disc in the centre of the hole, a 'star' collector, and a 'ladder' collector that produced a variety of different scaffold architectures. In the manufacture of PCL scaffolds, a solvent system of chloroform/dimethylformamide or cholorform/methanol has been a route of choice, due to the conductivity of the solution, although the templating of other synthetic polymers such as poly(D,L-lactide- co-glycolide) has also been reported (Zhou et al., 2010).

A disadvantage of many templating techniques is that the structures remain relatively two-dimensional and do not truly biomimic the three-dimensional nature of physiological tissue such as bone. Various methods to produce multilayered structures have been explored, including stacking templated electrospun webs and fusing them using a hot pressing process (Dou et al., 2011). Collagen glue has also been used as an adhesive between stacked layers of scaffold (McCullen et al., 2010). Although such scaffolds are three-dimensional in terms of macroscopic thickness, the fibres remain orientated in the x–y directions and the pore structure reflects this. When multiple layers of web are stacked directly on top of each other, a reduction in the overall pore size can result, which is undesirable.

Three-dimensional collectors can be used to collect electrospun fibres in architectures similar to that of a cotton ball. Blakeney et al. (2011) reported the application of a spherical foam dish with stainless steel probes radiating from its centre. This collector was used to deposit PCL fibres and produced a low-density, three-dimensional structure. Fibre diameters of 500 nm and pore sizes between 2 and 5 mm were obtained. Despite the original design of this collector, the pore size remained far below the suggested minimum pore size required for bone engineering of 100 nm.

Direct-write electrospinning is also emerging as a means for introducing biomimetic features. The original direct-write technology allowed structures to be built directly, without the use of masks enabling the rapid prototyping of complex architectures (Chrisey, 2000). Combining the technology with electrospinning facilitates the production of complex fibrous architectures by additive manufacture. Both solution and melt electrospinning have been demonstrated to be compatible with direct writing. To enable direct writing, the electrospinning process is modified so that the fibres can be collected while the fibre deposition path is still straight. At the point of fibre contact with the collector, the majority of the solvent should either have been removed or the melt solidified, so that the fibres retain their incoming morphology. To enable the direct writing in this way, a fast moving lateral collector platform is required. Polymer melts can be drawn over a larger distance while remaining on a straight trajectory, as compared to their solution-spun counterparts. This allows the tip-to-collector distance to be greater, resulting in a longer time period for the molten polymer to solidify. In order to be able to write with a continuous line, the translational speed of the collector must match the jet speed at impact (up to 1 m/min in some instances). Consideration must also be given to the turning speed/dwell time during the writing of complex patterns and incorporated into the original programmed design. These patterned webs can be repeatedly manufactured on top of each other to produce 3D structures with up to 1 cm thickness (Brown et al., 2012). To date, PCL has been frequently used because of its wide range of processing temperatures, allowing the production of fibres with a range of diameters between 19 and 28 mm (Brown et al., 2012). The resulting architectures are often geometric in design and made of multiple layers of large diameter fibres.

Solution-based electrospun direct writing requires a similar set-up, but because of the time required for the solvent to be flashed off, a larger tip-to-collector distance is required. The set-up used by Lee et al. (2012) comprised a dielectric thin plate that was capable of lateral movement independent of the sharp-pin ground electrode. Nanofibres were collected as dense depositions in a defined area (the pin electrode), and writing was accomplished by movement of the collector surface. To add stability and three-dimensionality, multiple layers of nanofibres must be written on top of each other. Various patterns, such as lines, points, lattices, and grids, can be produced. Polymer selection is limited to those that can be spun successfully in volatile solvents such as chloroform, ensuring that the majority of the solvent is evaporated prior to the fibre depositing upon the collector. Although greater control of fibre deposition is achieved, the architectures are still relatively 2D in construction.

Electrospinning into a liquid coagulating bath is a combination of electrospinning and conventional wet spinning. Wet electrospinning has received considerable attention due to its potential for decreasing fibre packing density and increasing the pore size in scaffolds. By using a solvent with a relatively low surface tension, fibres do not float on the surface and can disperse in the solvent, allowing architectures with a more pronounced three-dimensional structure to be formed (Yokoyama et al., 2009). Further increases in pore size can be obtained by incorporating porogens such as sodium chloride into the spinning bath (Gang et al., 2012). Synthetic polymers such as poly(glycolic acid) have been spun in solutions of 1,1,1,3,3,3-hexafluoro-2-propanol and collected in a bath of tertiary-butyl alcohol. The resulting architectures contained fibres with diameters ranging from 200 to 1400 nm. PCL has been spun into ethanol, which prevented the electrospun fibres from packing densely on top of each other (Yang et al., 2012). To prevent the architecture from collapsing when the solvent is removed, freeze-drying is required after spinning.

The versatility of electrospinning as a technique for bone scaffold production is considerable, particularly when it is combined with techniques that enable better control of the fibrous architecture. Unfortunately, although a great variety of structural features can be introduced, the results cannot be described as truly biomimetic. Alternative nonwoven technologies are likely to be required to overcome some of the inherent limitations of electrospinning, and there is substantial scope given the range of processes that are already commercially available. The challenge remains to engineer improved electrospun 3D architectures with both pore sizes and mechanical properties that are properly matched to clinical needs.

## 4. Considerations for surgical implantation of nonwoven scaffolds

To meet the specific clinical requirement, the researcher must consider many factors when designing and fabricating nonwoven scaffolds. The general requirements of biomaterials and scaffold architectures have already been discussed; however, these may need to be tailored, depending on the requirements of the specific bone defect. Common surgical situations that would be suitable for reconstruction using a nonwoven scaffold are listed in Table 1. Depending on the defect type, nonwoven scaffolds can be applied to the defect site in different ways. For a fracture model, the nonwoven scaffold may be used as a membrane/bandage to cover the fracture and/or bone defect area, whereas, for large defect sites, the nonwoven scaffold may be used as filler/ gauze, as shown in Figure 3. It is important to identify the intended use of nonwoven scaffolds and to incorporate appropriate features during the scaffold development phase. The final scaffold structure must be user-friendly and should be sterile and supplied in a range of different sizes. It must also be ready to use and easy to handle by the surgeon.

**Table 1.** Common surgical situations suitable for reconstruction using a nonwoven scaffold.

| Bone defect type | Examples of surgical situations | Reference |
|---|---|---|
| Fracture nonunion: small bone defect after surgical procedure | Removal of benign tumour: unicameral bone cysts, periapical cyst, or tumour resection. | Khira and Badawy (2013), Di Stefano et al. (2012), and Gentile et al. (2013) |
| Segmental bone defect | Due to trauma or surgery | Liu et al. (2013a), Gruber et al. (2013), Horner et al. (2010), and Udehiya et al. (2013) |
| Flat bone | Calvaria defect | Liu et al. (2013b), Pelegrine et al. (2013), and Cooper et al. (2010) |

Consideration must also be given to the method of fixing the scaffold in the defect site. Not all nonwoven scaffolds require additional fixation because they can be pushed into place to fill the defect area. Push-fit nonwovens scaffolds must have sufficient recovery post-implantation to be able to mould to the contours of the defect site, thereby remaining in situ. Biological glue or surgical adhesives may be helpful for keeping the scaffold in place (Kukleta et al., 2012). In a flat bone defect such as the calvaria defect, nonwoven scaffolds have the advantage of being able to be cut to the shape of the defect (Figure 4a). A well-fitting scaffold may not need any fixation, however, fixing may be needed if the bone defect is large. In such as case, a temporary protective layer/tissue flap may be required to cover the defect area before new bone is formed. Similarly, for nonunion fractures, the nonwoven scaffold may be kept in place by the surrounding tissues or with biological glue or surgical adhesives. In the case of spinal fusion, the nonwoven scaffold may be used as filler/ gauze and/ or strips to fill the defect area (Figure 4b), kept in place by the surrounding tissues.

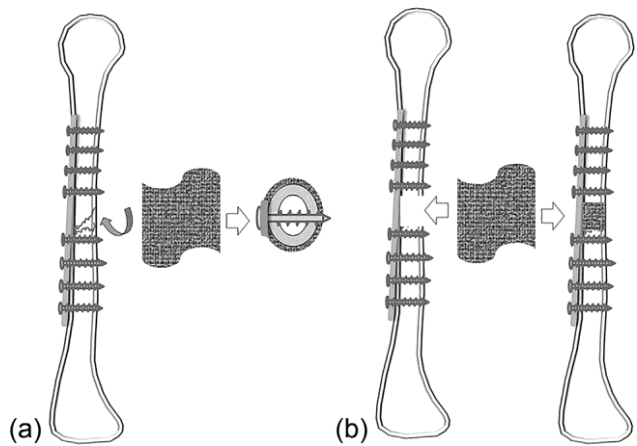

**Figure 3.** Possible use of nonwoven scaffolds for fracture repair/ bone tissue regeneration (in combination with/without cells and/or growth factors). (a) Fracture model: the nonwoven scaffold may be used as membrane/bandage to cover the fracture and/or bone defect area; (b) bone defect model: the nonwoven scaffold may be used as filler/gauze to fill the bone defect area.

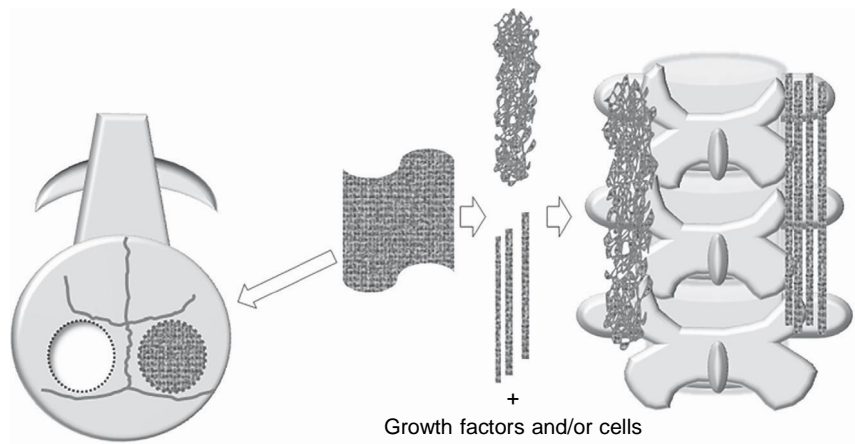

**Figure 4.** Possible application of nonwoven scaffolds in non-weight-bearing areas (in combination with/ without cells and/or growth factors). (a) Calvarial defect model: the nonwoven scaffold may be used as membrane/ gauze to fill the bone defect area; (b) spinal fusion model: the nonwoven scaffold may be used as filler/ gauze/ strips to fill areas that are required for spinal fusion.

## 5. Future trends

For nonwoven scaffolds to meet clinical requirements for bone repair and regeneration, they must be able to successfully biomimic aspects of the native tissue, and they must be sufficiently robust for surgical use. Furthermore, they need to be environmentally stable, reproducible in structure, and economically produced from appropriate biomaterials. Meeting these requirements using existing manufacturing techniques, such as electrospinning, is challenging, but the technology is rapidly evolving, and other multistage nonwoven manufacturing methods have yet to be fully explored as alternative production routes for bone scaffolds. Developments in biomaterial science are still required in order to provide fibre-forming materials that are compatible with large scale manufacture while satisfying all performance requirements. Biomaterials such as those based upon functionalised collagen have significant potential for the engineering of nonwoven scaffolds that address both biological and mechanical property requirements. The role of clinicians in guiding tissue scaffold design will continue to be fundamental to the development of functional scaffolds that meet individual patient needs. Better matching of scaffold properties to individual needs is essential if clinical outcomes are to improve, and this can increase the demand for the customisation of nonwoven scaffolds during manufacture and at the bedside. These developments have the potential to make a major impact on current scaffold manufacturing procedures, as well as the development of the regulatory framework.

## Acknowledgements

The authors are supported by the Leeds Centre of Excellence in Medical Engineering funded by the Wellcome Trust and EPSRC, WT088908/z/09/z.